\pdfoutput=1

\documentclass[11pt,twoside,a4paper,cmspaper,final,collab]{cms-tdr}
\def\svnVersion{71540:71544}\def\cmsCernNo{2011-119}\def\cmsCernDate{\today}\def\cmsMessage{Submitted to Physics Letters B}

\begin{document}\cmsNoteHeader{EXO-11-015}

\hyphenation{had-ron-i-za-tion}
\hyphenation{cal-or-i-me-ter}
\hyphenation{de-vices}

\RCS$Revision: 71544 $
\RCS$HeadURL: svn+ssh://alverson@svn.cern.ch/reps/tdr2/papers/EXO-11-015/trunk/EXO-11-015.tex $
\RCS$Id: EXO-11-015.tex 71544 2011-07-24 09:57:29Z alverson $
\newlength{\cmsFigWidth}
\ifthenelse{\boolean{cms@external}}{\setlength{\cmsFigWidth}{0.9\columnwidth}}{\setlength{\cmsFigWidth}{0.55\textwidth}}
\cmsNoteHeader{EXO-11-015} 
\title{Search for Resonances in the Dijet Mass Spectrum from 7~TeV pp Collisions at CMS}

\date{\today}

\abstract{
A search for narrow resonances with a mass of at least 1 TeV in the dijet mass spectrum
is performed using  pp collisions at $\sqrt{s}=7$~TeV corresponding to an integrated luminosity of 1
fb$^{-1}$, collected
by the CMS experiment at the LHC. No resonances are observed. Upper limits at the 95\% confidence level are presented on the product of the resonance cross
section, branching fraction into dijets, and acceptance, separately for decays into
quark-quark, quark-gluon, and gluon-gluon pairs. The data
exclude new particles predicted in the following models at the
95\% confidence level: string resonances with mass less than $4.00$~TeV, $\mbox{E}_6$ diquarks with mass less than $3.52$~TeV,
excited quarks with mass less than $2.49$~TeV, axigluons and colorons with mass less than $2.47$~TeV,
and $\mbox{W}^\prime$ bosons with mass less than $1.51$~TeV.}

\hypersetup{
pdfauthor={CMS Collaboration},
pdftitle={Search for Resonances in the Dijet Mass Spectrum from 7 TeV pp Collisions at CMS},%
pdfsubject={CMS},%
pdfkeywords={CMS, physics}}

\maketitle 

The Large Hadron Collider (LHC) has recently delivered an integrated
luminosity in excess of 1 fb$^{-1}$ at a centre-of-mass energy $\sqrt{s}=7$~TeV.
This extends considerably the search territory for new physics in events containing
jets.
In this Letter we report a search for narrow
resonances in the dijet mass spectrum, performed with the Compact Muon Solenoid (CMS)
detector~\cite{refCMS}, with sensitivity exceeding that of our previous search~\cite{Khachatryan:2010jd}.
Proton-proton collisions produce two or more energetic jets when the constituent partons are
scattered with large transverse momenta, $p_T$.  The invariant mass spectrum of the two jets with
largest $p_T$ (dijets) is predicted to fall steeply and smoothly by quantum
chromodynamics (QCD). Many extensions of the standard model predict the existence of new massive objects
that couple to quarks (q) and gluons (g), and result in resonances in the
dijet mass spectrum.

We apply the results of this generic search to the following specific models
of narrow $s$-channel dijet resonances:
\begin{itemize}
\item String resonances (S), which are Regge excitations of quarks and gluons in string
theory and decay predominantly to qg~\cite{Anchordoqui:2008di,Cullen:2000ef}.
\item Scalar diquarks (D), which decay to qq and $\bar{\mbox{q}}\bar{\mbox{q}}$, predicted by a grand unified theory based on
the $\mbox{E}_6$ gauge symmetry group~\cite{ref_diquark}.
\item Mass-degenerate excited quarks ($\mbox{q}^*$), which decay to
qg, predicted if quarks are composite objects~\cite{ref_qstar,Baur:1989kv}; the compositeness scale is set to be equal to the mass of the excited quark.
\item Axial-vector particles called axigluons (A), which decay to  $\mbox{q}\bar{\mbox{q}}$,
predicted in a model where the symmetry group $\hbox{SU(3)}$ of QCD is replaced by the
chiral symmetry ${\hbox{SU(3)}_{\rm L}} \times {\hbox{SU(3)}_{\rm R}}$~\cite{ref_axi}.
\item Color-octet colorons (C), also decaying to  $\mbox{q}\bar{\mbox{q}}$, predicted by the flavour-universal
coloron model, embedding the $\hbox{SU(3)}$ symmetry of QCD in
a larger gauge group~\cite{ref_coloron}.
\item New
gauge bosons ($\mbox{W}^\prime$ and $\mbox{Z}^\prime$), which decay to $\mbox{q}\bar{\mbox{q}}$,
predicted by models that include new gauge symmetries~\cite{ref_gauge};
the $\mbox{W}^\prime$ and $\mbox{Z}^\prime$ bosons
are assumed to have standard model couplings.
\item Randall-Sundrum (RS) gravitons (G), which decay to $\mbox{q}\bar{\mbox{q}}$ and gg, predicted in the RS model
of extra dimensions~\cite{ref_rsg}; the value of the dimensionless coupling
$\kappa/\overline{M}_{\rm Pl}$ is chosen to be 0.1.
\end{itemize}

The central feature of the CMS apparatus is a superconducting solenoid of 6 m
internal diameter providing an axial field of 3.8 tesla.
Within the field volume at central values of pseudorapidity $\eta$ are the silicon pixel and strip tracker ($|\eta|<2.4$) and the barrel and
endcap calorimeters ($|\eta|<3$): a lead tungstate crystal electromagnetic calorimeter (ECAL) and
a brass/scintillator hadronic calorimeter (HCAL).
An iron/quartz-fiber calorimeter is located in the forward region ($3<|\eta|<5$), outside the field volume.
For triggering purposes and to facilitate jet reconstruction, the ECAL and HCAL
cells are grouped into towers projecting radially outward from the
centre of the detector.
The energy deposits measured in the ECAL and the
HCAL within each projective tower are summed to find the calorimeter tower energy.
A more detailed description of the CMS experiment can be found
elsewhere~\cite{refCMS}.

The CMS coordinate system has the origin at the center of the detector.
The $z$-axis points along the direction of the anticlockwise beam, with the transverse
plane perpendicular to the beam; $\phi$ is the azimuthal
angle, $\theta$ is the polar angle, and the pseudorapidity is defined as
$\eta \equiv -\ln(\tan[\theta/2])$.

The integrated luminosity of the data sample selected for this
analysis is $1.01 \pm 0.06$~fb$^{-1}$.
Events are recorded using a two-tier trigger system. Objects satisfying the
requirements at the first level (L1) are passed to the High Level Trigger (HLT).
The sample was collected with a multijet trigger at the HLT, which is based on the sum of
the transverse energies of all jets in the event with $p_T$ above 40 GeV.
The trigger selects events with a total transverse energy of jets in the
HLT exceeding 550~GeV.
The trigger efficiency is measured from the data
to be larger than 99.9\% for dijet masses above 838~GeV.

To remove possible instrumental and non-collision
backgrounds in the selected sample, jets are required to pass
identification criteria that are fully efficient for signal~\cite{JME-10-003-PAS}.
Events are
required to have a reconstructed primary vertex within the range $|z|<24$ cm.

We consider two types of standard jets with different inputs: particle-flow jets,
which we use for the search, and calorimeter jets, which we use as a check.
The particle-flow algorithm~\cite{PFT-09-001-PAS} reconstructs all stable particles
in an event by combining information from all
subdetectors. The algorithm categorizes all particles
into the following five types: muons, electrons, photons, charged hadrons, and neutral
hadrons. Particle-flow jets use reconstructed particles as input to the jet
reconstruction algorithm, while calorimeter jets use calorimeter energy deposits as
the input.

The reconstructed jet energy $E$ is defined as the scalar sum of the energies
of the constituents of the jet, and the jet momentum $\vec{p}$ is the corresponding vector sum of
the momenta of the inputs. The jet transverse momentum $p_T$ is the component
of $\vec{p}$ perpendicular to the beam.
The values of $E$ and $\vec{p}$ of a reconstructed jet are corrected for the
response of the detector to a generated jet, using Monte Carlo
simulations, test beam results, and collision data~\cite{Collaboration:2011ds}.

This analysis combines particle-flow jets reconstructed with the anti-$k_T$
algorithm~\cite{1126-6708-2008-04-063} into "wide jets", which we use to measure
the mass spectrum and search for new physics.  Wide jets are the result of a
radiation recovery algorithm for dijets, inspired by recent jet-grooming
algorithms~\cite{Krohn:2009th,Abdesselam:2010pt}.
The partons from the decay of heavy objects can radiate additional partons,
which are often produced at a large angle with respect to the original parton direction and
thus are clustered into a separate
jet by the anti-$k_T$ jet-clustering algorithm. Wide jets collect more of this final-state
radiation and therefore improve the mass resolution for dijet resonances.
First, we reconstruct AK5 and AK7 jets using the anti-$k_T$ algorithm
with distance parameters $R=0.5$ and $R=0.7$, respectively, which are the
two standard choices we support for analysis at CMS. In our previous
search~\cite{Khachatryan:2010jd} we used AK7 jets, since they have a larger distance
parameter than AK5 jets and capture more radiation. Here we introduce wide jets reconstructed from AK5 jets to
produce a wider jet than AK7.  We correct the AK5 jet energy and select the two AK5 jets with
the highest $p_T$ in the event (leading AK5 jets). Then we add the Lorentz vectors of
all other AK5 jets with $p_T>10$ GeV and $|\eta|<2.5$ to the closest AK5 leading jet, if within
$\Delta R=\sqrt{(\Delta\eta)^2 + (\Delta\phi)^2}<1.1$, to obtain the two leading wide jets.
The parameter $\Delta R$ sets the maximum size of the wide jet.

The dijet system is composed of the two leading jets.
We require that the pseudorapidity separation $\Delta\eta$ of the two leading jets satisfy
$|\Delta\eta|<1.3$, and that both jets be in the region
$|\eta|<2.5$. These $\Delta\eta$ and $\eta$ requirements maximize the search sensitivity for
isotropic decays of dijet resonances in the presence of QCD background.
The dijet mass is given by  $m=\sqrt{(E_1 + E_2)^2 - (\vec{p}_1 + \vec{p}_2)^2}$.
We select events with $m>838$~GeV without any requirements on
the $p_T$ of the leading jet.

\begin{figure}[htbp]
  \begin{center}
   \includegraphics[width=\cmsFigWidth]{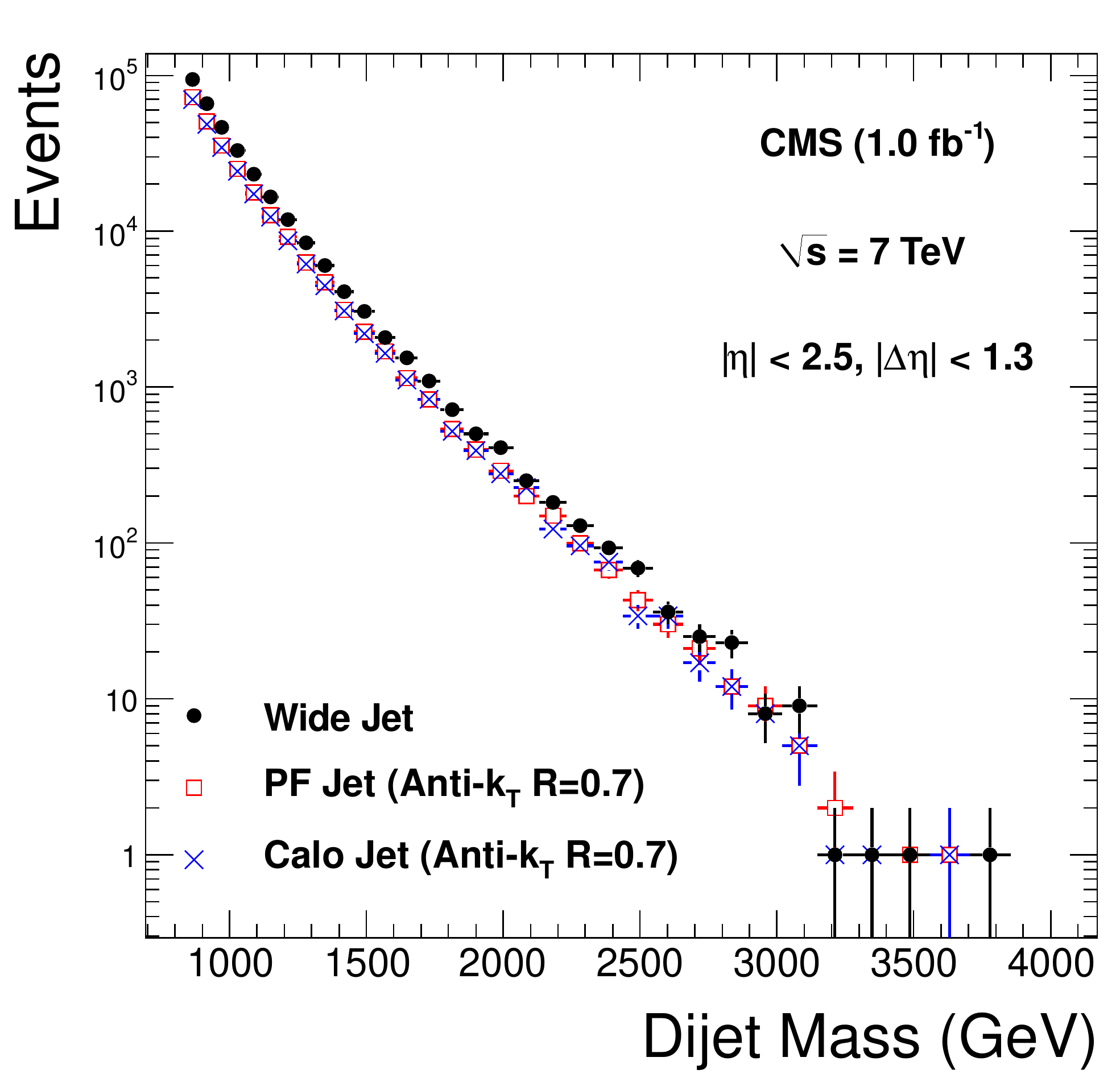}
    \caption{ The number of events observed versus dijet mass for wide jets (solid circles),
particle flow AK7 jets (open boxes), and calorimeter AK7 jets (X symbols).}
    \label{CaloAndPF}
  \end{center}
\end{figure}

 The number of events as a function of dijet mass is shown in
Fig.~\ref{CaloAndPF} for both calorimeter and particle-flow AK7 jets; the
observed rates agree. Figure~\ref{CaloAndPF} also shows that the observed wide jet
dijet mass distribution is shifted to higher mass because wide jets collect
more energy.

\begin{figure}[hbtp]
  \begin{center}
    \includegraphics[width=\cmsFigWidth]{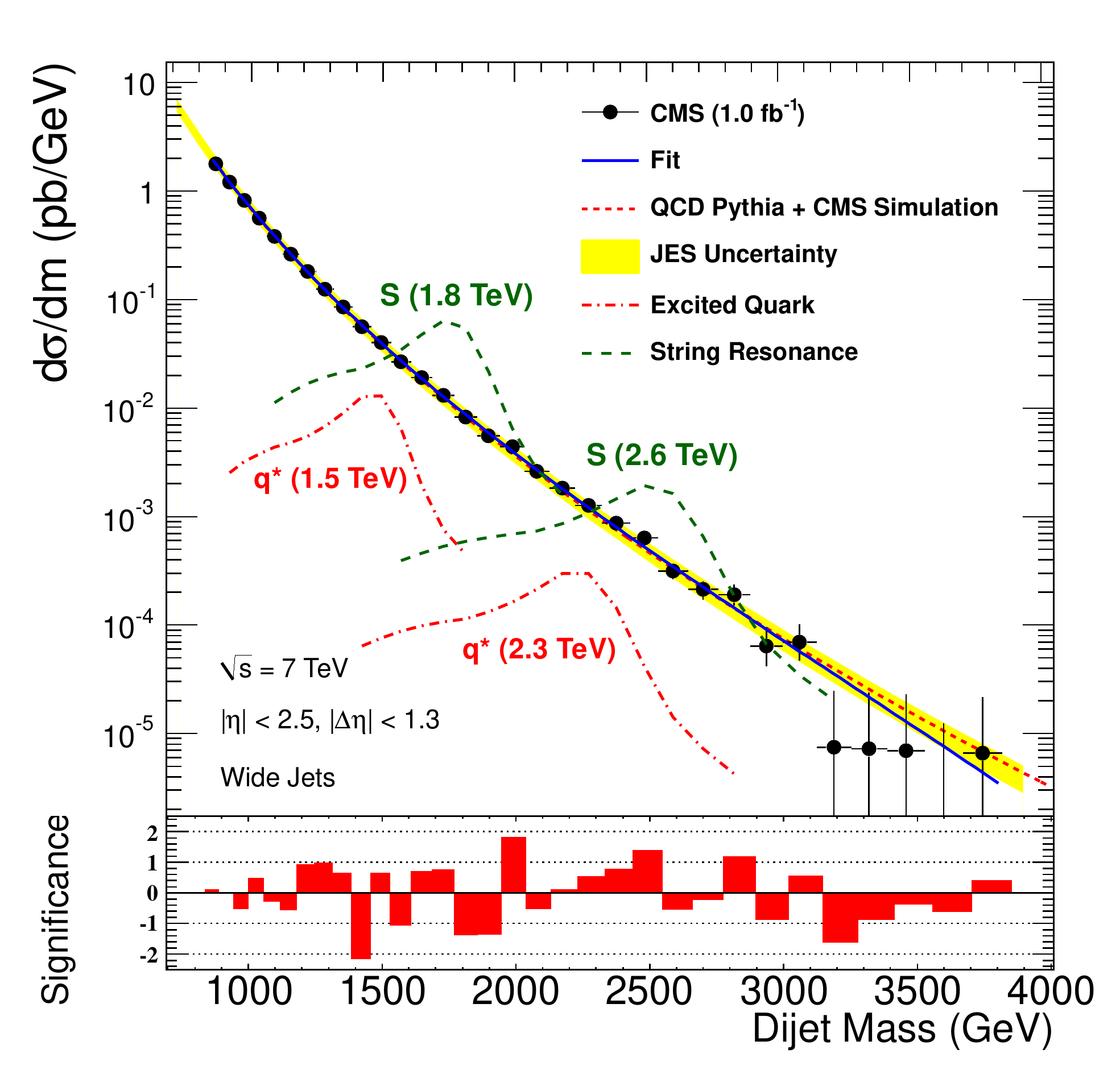}
    \hspace{1cm}
    \caption{
    Dijet mass spectrum from wide jets (points) compared to a smooth fit (solid)
    and to predictions~\cite{refPYTHIA} including detector simulation of QCD (short-dashed), excited quark signals (dot-dashed),
    and string resonance signals (long-dashed). The QCD prediction has been normalized to the data (see text).
    The error bars are statistical only.
     The shaded band shows the systematic uncertainty in the jet energy scale (JES).
     The bin-by-bin significance of the data-fit difference (see text) is shown at bottom.}
    \label{figData}
  \end{center}
\end{figure}

\begin{figure}[hbtp]
  \begin{center}
    \includegraphics[width=\cmsFigWidth]{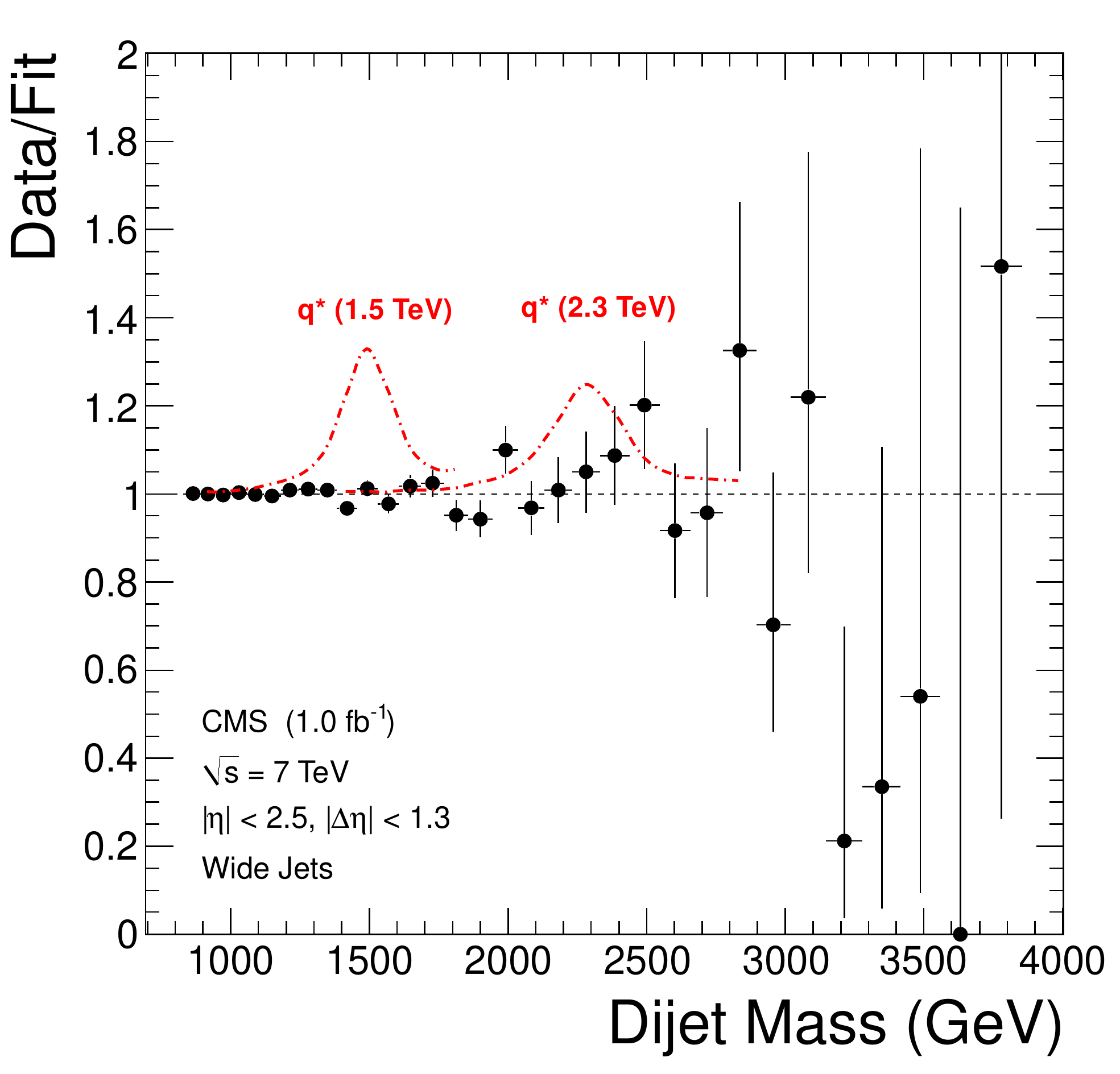}
    \hspace{1cm}
    \caption{Ratio (points) between the dijet mass data from wide jets and the smooth fit,
    compared to the simulated ratios for excited quark signals
(dot-dashed) in the CMS detector.
The error bars represent statistical uncertainties only.}
    \label{figDiff}
  \end{center}
\end{figure}

Figure~\ref{figData} presents the inclusive dijet mass
distribution for $\mbox{pp}\rightarrow$ 2 leading wide jets + X, where X can be
anything, including additional jets.
Wide jets are used and we plot the measured differential cross section as a function of dijet mass in
bins approximately equal to the dijet mass resolution~\cite{Khachatryan:2010jd}.
The data are compared to a QCD prediction from \PYTHIA~{\sc v6.424}~\cite{refPYTHIA}, which includes a simulation of the CMS
detector and the jet energy corrections.
The prediction uses a renormalization scale $\mu=p_T$ of the hard-scattered
partons and CTEQ6L1 parton distribution functions~\cite{refCTEQ},
and has been normalized to the data by multiplying the prediction by a factor of 1.33.
The shape of the \PYTHIA prediction agrees with the data within the jet energy scale uncertainty, which is the dominant
systematic uncertainty.
To test the smoothness of our measured cross section as a function of dijet mass, we
fit the following parameterization to the data:

\begin{equation}
\frac{{\rm d}\sigma}{{\rm d}m} =
\frac{P_{0} (1 - m/\sqrt{s})^{P_{1}}}{(m/\sqrt{s})^{P_{2} + P_{3} \ln{(m/\sqrt{s})}}}\ ,
\end{equation}

with four free parameters $P_0$, $P_1$, $P_2$, and $P_3$. This functional form
is used in previous searches~\cite{refCDFrun2,ATLAS2010,Khachatryan:2010jd,
Aad:2011aj} to describe both data and QCD predictions.
In Fig.~\ref{figData} we show the fit, which has a chi-squared ($\chi^2$) of 27.5 for 28 degrees of freedom, as
well as the bin-by-bin significance, defined as the difference between the data and the fit value, divided
by the statistical uncertainty of the data.
Figure~\ref{figDiff} displays the ratio of
the data to the fit.  The data are well described by
the smooth parameterization.

\begin{figure}[hbtp]
  \begin{center}
    \includegraphics[width=\cmsFigWidth]{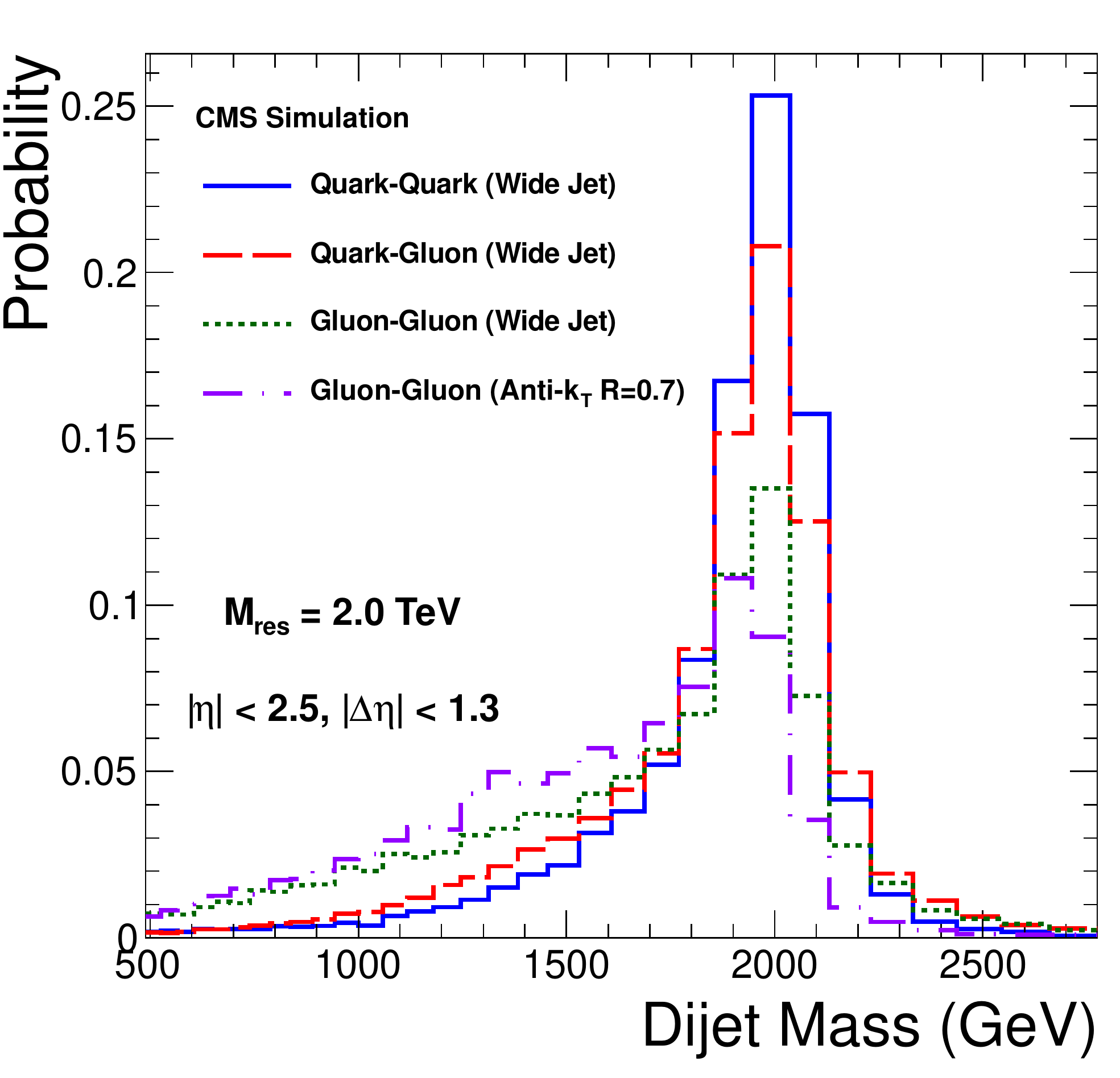}
    \hspace{1cm}
    \caption{ Simulation of the expected dijet mass distributions
in the CMS detector from a narrow 2.0~TeV
resonance of type quark-quark (solid), quark-gluon (dashed),
and gluon-gluon (dotted) using wide jet reconstruction, and of type gluon-gluon
using anti-$k_T$ jet reconstruction with distance parameter $R=0.7$ (dot-dash).}
    \label{figResonance}
  \end{center}
\end{figure}

We search for narrow resonances, for which the
natural resonance width is small compared to the CMS dijet mass
resolution. Figures~\ref{figData} and \ref{figDiff} present the
predicted dijet mass distribution for excited quarks using
\PYTHIA~{\sc v6.424} and the CMS detector simulation.
The predicted mass distributions have a Gaussian core coming from
the jet energy resolution and a tail towards lower mass from QCD radiation.
This can be seen in Fig.~\ref{figResonance}, which shows examples of the predicted dijet mass distribution of
resonances from
three different parton pairings: $\mbox{q}\bar{\mbox{q}}$ (or qq) resonances
from the process $\mbox{G} \rightarrow \mbox{q}\bar{\mbox{q}}$~\cite{ref_rsg},
qg resonances from $\mbox{q}^* \rightarrow \mbox{qg}$~\cite{ref_qstar},
and gg resonances from $\mbox{G} \rightarrow \mbox{gg}$~\cite{ref_rsg}.
The increase of the
width of the measured mass shape and the shift of the mass
distribution towards lower masses are enhanced when the number of
gluons in the final state is larger, because QCD radiation is larger
for gluons than for quarks.
The distributions in Fig.~\ref{figResonance} are generically valid for other resonances with the
same parton content and with a natural width small compared to the dijet mass
resolution, and are examples of the shapes we use to set limits on dijet resonances.
Wide-jet reconstruction gives a little better resolution than AK7-jet reconstruction,
as shown in Fig.~\ref{figResonance} for gg resonances.
 There is no indication
of narrow resonances in our data, as shown in Figs.~\ref{figData} and ~\ref{figDiff}.

\begin{table*}[htbp]
\begin{center}
\caption{
The observed upper limits at the 95\% CL on $\sigma\times B\times A$,
as a function of the new-particle mass, for narrow resonances decaying to dijets
with partons of type quark-quark (qq), quark-gluon (qg), and gluon-gluon (gg).
The limits apply to the kinematic range where the two jets have pseudorapidity
$|\eta|<2.5$ and $|\Delta\eta|<1.3$.
}
\label{tab_limit}
\vspace*{1cm}
\begin{tabular}{|c|l|l|l||c|l|l|l|}\hline
Mass   &  \multicolumn{3}{c||}{Upper limit (pb)} & Mass   &
\multicolumn{3}{c|}{Upper limit (pb)}\\
 (TeV) &  qq & qg & gg &  (TeV) &  qq & qg & gg\\ \hline
1.0 & 1.098 & 1.245 & 1.851 & 2.6 & 0.0619 & 0.0749 & 0.1011 \\
1.1 & 0.777 & 0.909 & 1.374 & 2.7 & 0.0470 & 0.0577 & 0.0781 \\
1.2 & 0.662 & 0.732 & 1.079 & 2.8 & 0.0356 & 0.0440 & 0.0601 \\
1.3 & 0.486 & 0.535 & 0.803 & 2.9 & 0.0274 & 0.0336 & 0.0455 \\
1.4 & 0.284 & 0.332 & 0.518 & 3.0 & 0.0210 & 0.0257 & 0.0348 \\
1.5 & 0.231 & 0.265 & 0.395 & 3.1 & 0.0158 & 0.0196 & 0.0268 \\
1.6 & 0.201 & 0.226 & 0.326 & 3.2 & 0.0118 & 0.0151 & 0.0209 \\
1.7 & 0.168 & 0.190 & 0.280 & 3.3 & 0.0091 & 0.0117 & 0.0163 \\
1.8 & 0.115 & 0.138 & 0.207 & 3.4 & 0.0075 & 0.0097 & 0.0136 \\
1.9 & 0.113 & 0.131 & 0.183 & 3.5 & 0.0065 & 0.0085 & 0.0117 \\
2.0 & 0.121 & 0.140 & 0.193 & 3.6 & 0.0061 & 0.0078 & 0.0105 \\
2.1 & 0.108 & 0.130 & 0.183 & 3.7 & 0.0059 & 0.0074 & 0.0096 \\
2.2 & 0.093 & 0.115 & 0.160 & 3.8 & 0.0057 & 0.0071 & 0.0091 \\
2.3 & 0.089 & 0.108 & 0.148 & 3.9 & 0.0055 & 0.0068 & 0.0084 \\
2.4 & 0.085 & 0.102 & 0.138 & 4.0 & 0.0052 & 0.0064 & 0.0080 \\
2.5 & 0.077 & 0.092 & 0.125 & 4.1 & 0.0048 & 0.0060 & 0.0073 \\
\hline
\end{tabular}
\end{center}
\end{table*}

We use the dijet mass data from wide jets, the background (QCD) parameterization, and the
dijet resonance shapes to set specific limits on new particles decaying to
the parton pairs qq (or $\mbox{q}\bar{\mbox{q}}$), qg, and gg.
The dominant sources of systematic uncertainty are the jet energy scale (2.2\%), the jet energy
resolution (10\%), the integrated luminosity (6\%), and the statistical uncertainty on the background
parameterization, which are all considered nuisance parameters.
For setting upper limits we use a Bayesian
formalism with a uniform prior for the signal cross section~\cite{Nakamura:2010zzi}.
To incorporate systematic uncertainties we use a fully Bayesian treatment,
integrating the likelihood over these nuisance parameters.
We calculate the posterior probability density as a function of resonance cross section independently at each value of the resonance mass.
Table~\ref{tab_limit} lists the generic upper limits at the 95\% confidence
level (CL) on
$\sigma\times B\times A$, i.e. the product of
the cross section ($\sigma$), the branching fraction ($B$), and the acceptance ($A$), for the
kinematic requirements $|\Delta\eta|<1.3$ and $|\eta|<2.5$,
for qq, qg, and gg resonances. The acceptance for isotropic decays
is $A\approx 0.6$ independent of resonance mass.  The observed upper limits in Table~\ref{tab_limit}
can be compared to predictions of $\sigma\times B\times A$
at the parton level, without any detector simulation, in order to determine mass
limits on new particles.
In addition to these observed upper limits,
we also calculate the expected upper limits using pseudo-experiments:
searches conducted on random samples of events generated from our smooth background parameterization.
The use of wide jets instead of AK7 jets improves the expected upper
limits on the resonance cross section by roughly 20\% for gg, 10\% for qg, and 5\%
for qq resonances.
\begin{figure}[hbtp]
  \begin{center}
    \includegraphics[width=\cmsFigWidth]{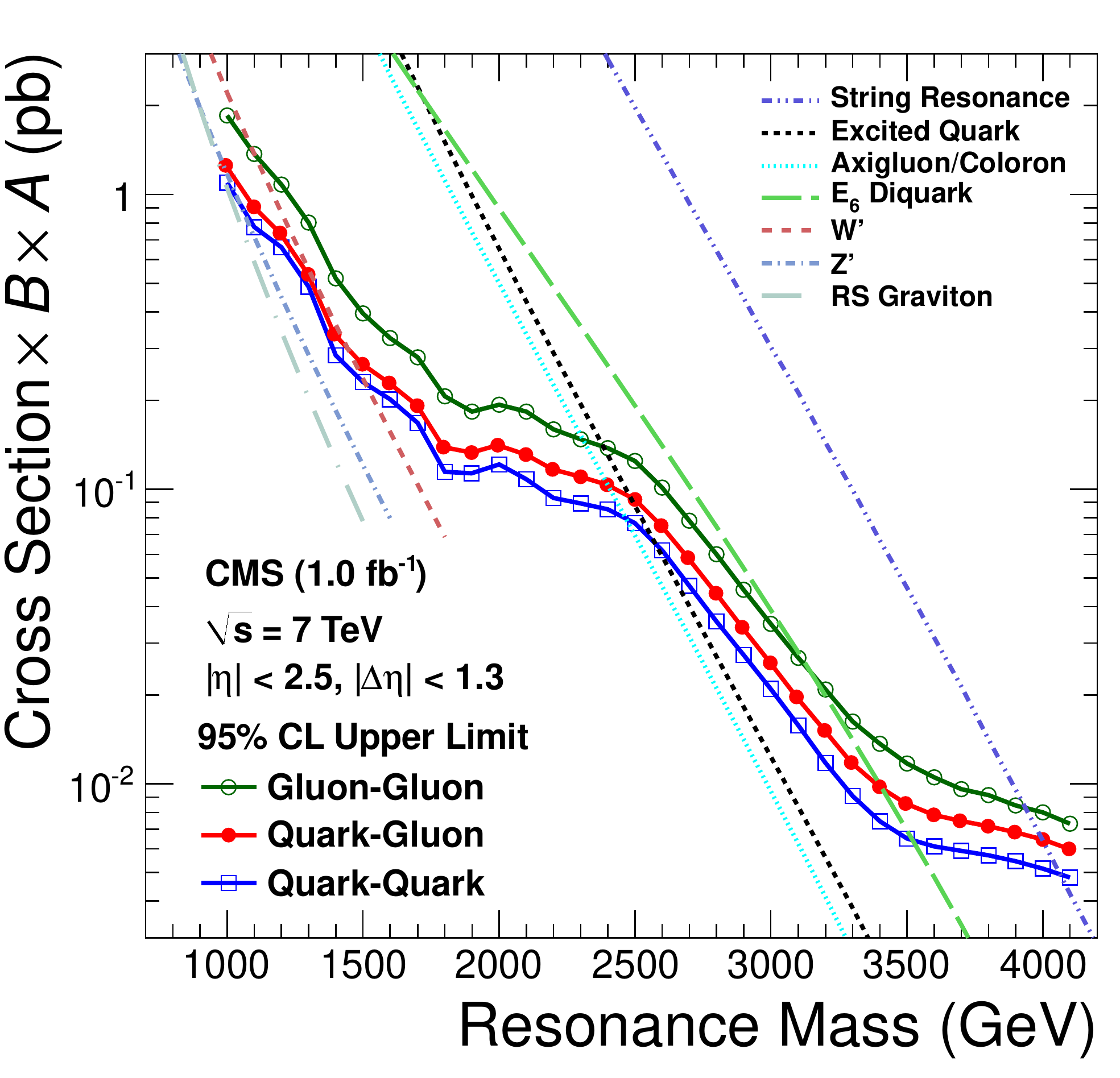}
    \hspace{1cm}
    \caption{The 95\% CL upper limits on $\sigma\times B\times A$
for dijet resonances of type gluon-gluon
(open circles), quark-gluon (solid circles), and quark-quark (open boxes),
compared to
theoretical predictions for string resonances~\cite{Anchordoqui:2008di}, $\mbox{E}_6$ diquarks~\cite{ref_diquark},
excited quarks~\cite{ref_qstar}, axigluons~\cite{ref_axi},
colorons~\cite{ref_coloron},
new gauge bosons $\mbox{W}^{\prime}$ and $\mbox{Z}^{\prime}$~\cite{ref_gauge},
and Randall-Sundrum gravitons~\cite{ref_rsg}.}
    \label{figLimit}
  \end{center}
\end{figure}

\begin{figure}[hbtp]
  \begin{center}
    \includegraphics[width=\cmsFigWidth]{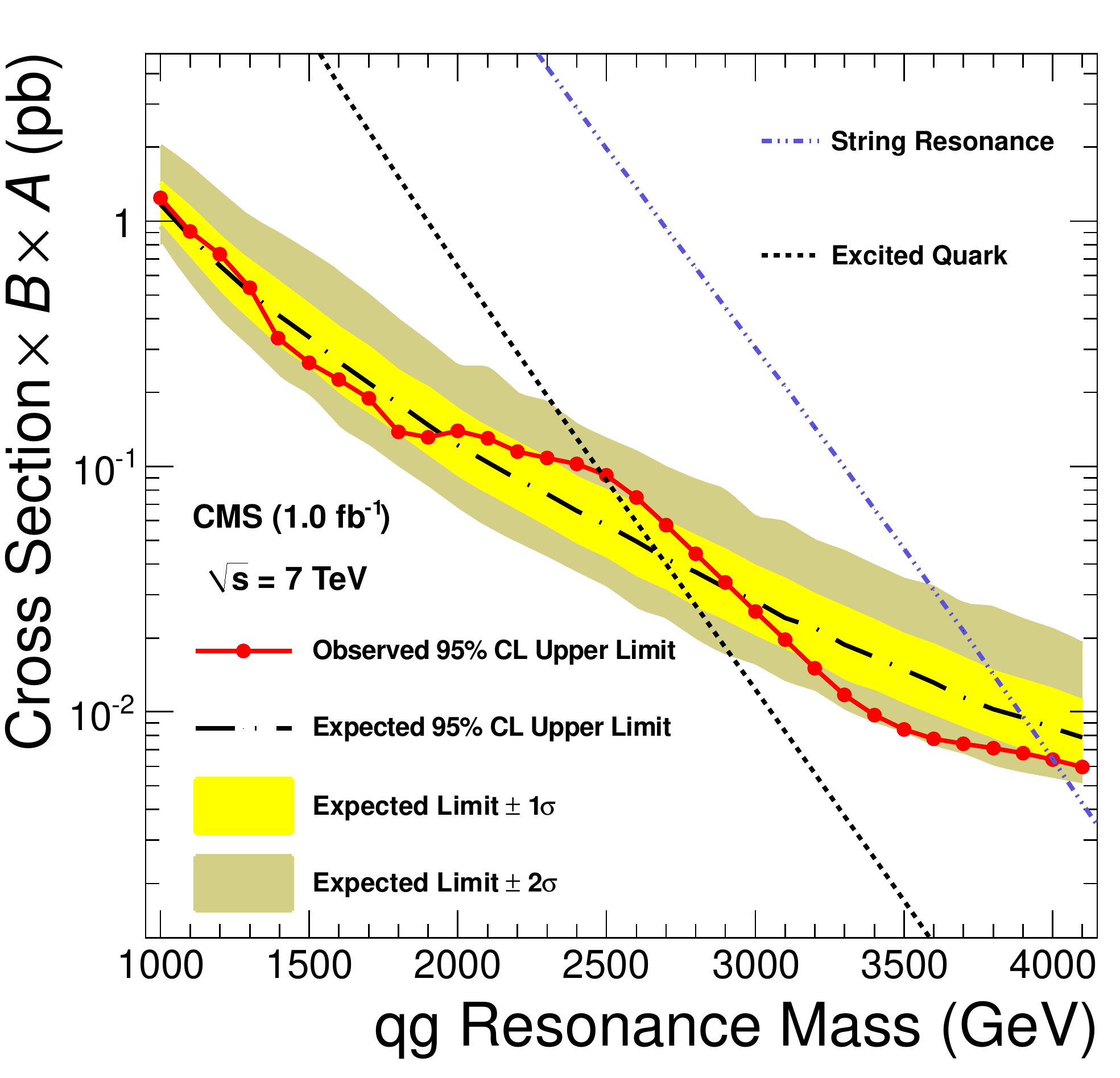}
    \hspace{1cm}
    \caption{The observed 95\% CL upper limits on $\sigma\times B \times A$
for quark-gluon dijet resonances (points) are compared to the expected limits
(dot-dash) and their expected statistical variation at the 1$\sigma$ and
2$\sigma$ levels (shaded bands) and are also compared to
the theoretical predictions for string resonances~\cite{Anchordoqui:2008di} and
excited quarks~\cite{ref_qstar}.}
    \label{figExpectedLimitqg}
  \end{center}
\end{figure}

\begin{table}[htbp]
\begin{center}
\caption{
For each model we list the observed and expected
upper values of the excluded mass range at 95\% CL. The lower value
of the excluded mass range from this search is 1 TeV.}
\label{tabMassLimit}
\vspace*{1cm}
\begin{tabular}{|c|c|c|}\hline
  Model            &  \multicolumn{2}{c|}{Excluded Mass (TeV)}\\
                   &  Observed    & Expected \\  \hline
String Resonances  &    4.00      & 3.90     \\
E$_6$ Diquarks     &    3.52      & 3.28     \\
Excited Quarks     &    2.49      & 2.68     \\
Axigluons/Colorons &    2.47      & 2.66     \\
W' Bosons          &    1.51      & 1.40     \\
\hline
\end{tabular}
\end{center}
\end{table}

In Fig.~\ref{figLimit} we compare the observed upper limits to the model predictions as a function of resonance mass.
The predictions are from lowest-order calculations~\cite{modelCalc} of the product $\sigma\times B\times A$
using CTEQ6L1 parton distributions~\cite{refCTEQ}. New particles are excluded at the 95\% CL in
mass regions for
which the theory curve lies above our upper limit for the appropriate pair of
partons.
We also determine the expected lower limit on the mass of each new particle by comparing the
expected cross section limits to the model predictions.
An example of the expected limits is shown in Fig.~\ref{figExpectedLimitqg} where for qg resonances we
compare the expected limits and their uncertainty bands to both observed limits and model predictions.
Our search starts at a resonance mass of 1 TeV, and all quoted exclusions start at this mass value.
The observed and expected limits are listed in Table~\ref{tabMassLimit}.
For string resonances the expected mass limit is $3.90$~TeV and we exclude
masses less than $4.00$~TeV; this extends our previous exclusion of
$0.5 < M(\mbox{S}) < 2.5$~TeV~\cite{Khachatryan:2010jd}.
For $\mbox{E}_6$ diquarks the expected limit is $3.28$~TeV and we
exclude masses less than $3.52$~TeV; this extends our previous exclusions
of $0.50<M(\mbox{D})<0.58$~TeV, $0.97 < M(\mbox{D}) < 1.08$~TeV, and
$1.45 < M(\mbox{D}) < 1.60$~TeV~\cite{Khachatryan:2010jd}.
For excited quarks the expected mass
limit is $2.68$ TeV and we exclude masses less than $2.49$~TeV; this extends
our previous exclusion of $0.5<M(\mbox{q}^*)<1.58$~TeV~\cite{Khachatryan:2010jd} and the
ATLAS exclusion of $0.6<M(\mbox{q}^*)<2.15$~TeV from the dijet mass spectrum~\cite{Aad:2011aj}.
For axigluons or colorons the
expected mass limit is $2.66$~TeV and we exclude masses less than
$2.47$~TeV; this extends our previous exclusions of $0.50<M(\mbox{A},\mbox{C})<1.17$~TeV and
$1.47<M(\mbox{A},\mbox{C})<1.52$~TeV~\cite{Khachatryan:2010jd}, and the ATLAS exclusion of
$0.6<M(\mbox{A},\mbox{C})<2.1$~TeV~\cite{Aad:2011aj}. For $\mbox{W}^\prime$ bosons the expected mass limit is
$1.40$~TeV and we exclude masses less than $1.51$~TeV; this extends
the CDF exclusion of $0.3 < M(\mbox{W}^\prime) < 0.8$~TeV from the dijet mass spectrum~\cite{refCDFrun2}.
We do not set any mass limits on $\mbox{Z}^\prime$ bosons and RS gravitons.
The systematic uncertainties included in this analysis reduce the excluded upper masses
by $0.03$~TeV or less for each type of new particle.

In summary, the dijet invariant mass distribution has been measured to be a smoothly falling
distribution, as expected within the standard model.
There is no evidence for new particle production. We present generic upper limits
on the product $\sigma\times B\times A$ that can be applied to any model
of dijet resonance production. We set specific mass limits on string resonances, $\mbox{E}_6$ diquarks, excited quarks,
axigluons, flavour-universal colorons, and $\mbox{W}^\prime$ bosons, all of which extend previous
exclusions from the dijet mass search technique.

\bigskip

We wish to congratulate our colleagues in the CERN accelerator departments for the excellent performance of the LHC machine.
We thank the technical and administrative staff at CERN and other CMS institutes, and acknowledge support from: FMSR (Austria); FNRS and FWO (Belgium); CNPq, CAPES, FAPERJ, and FAPESP (Brazil); MES (Bulgaria); CERN; CAS, MoST, and NSFC (China); COLCIENCIAS (Colombia); MSES (Croatia); RPF (Cyprus); Academy of Sciences and NICPB (Estonia); Academy of Finland, ME, and HIP (Finland); CEA and CNRS/IN2P3 (France); BMBF, DFG, and HGF (Germany); GSRT (Greece); OTKA and NKTH (Hungary); DAE and DST (India); IPM (Iran); SFI (Ireland); INFN (Italy); NRF and WCU (Korea); LAS (Lithuania); CINVESTAV, CONACYT, SEP, and UASLP-FAI (Mexico); PAEC (Pakistan); SCSR (Poland); FCT (Portugal); JINR (Armenia, Belarus, Georgia, Ukraine, Uzbekistan); MST and MAE (Russia); MSTD (Serbia); MICINN and CPAN (Spain); Swiss Funding Agencies (Switzerland); NSC (Taipei); TUBITAK and TAEK (Turkey); STFC (United Kingdom); DOE and NSF (USA).
We thank Can Kilic for calculations of the string resonance cross section.

\bibliography{auto_generated}   

\cleardoublepage \appendix\section{The CMS Collaboration \label{app:collab}}\begin{sloppypar}\hyphenpenalty=5000\widowpenalty=500\clubpenalty=5000\textbf{Yerevan Physics Institute,  Yerevan,  Armenia}\\*[0pt]
S.~Chatrchyan, V.~Khachatryan, A.M.~Sirunyan, A.~Tumasyan
\vskip\cmsinstskip
\textbf{Institut f\"{u}r Hochenergiephysik der OeAW,  Wien,  Austria}\\*[0pt]
W.~Adam, T.~Bergauer, M.~Dragicevic, J.~Er\"{o}, C.~Fabjan, M.~Friedl, R.~Fr\"{u}hwirth, V.M.~Ghete, J.~Hammer\cmsAuthorMark{1}, S.~H\"{a}nsel, M.~Hoch, N.~H\"{o}rmann, J.~Hrubec, M.~Jeitler, W.~Kiesenhofer, M.~Krammer, D.~Liko, I.~Mikulec, M.~Pernicka, B.~Rahbaran, H.~Rohringer, R.~Sch\"{o}fbeck, J.~Strauss, A.~Taurok, F.~Teischinger, C.~Trauner, P.~Wagner, W.~Waltenberger, G.~Walzel, E.~Widl, C.-E.~Wulz
% ...
\vskip\cmsinstskip
% ...
1:~~Also at CERN, European Organization for Nuclear Research, Geneva, Switzerland\\
% ...

\end{sloppypar}
\end{document}